# Controlling the Electronic Properties of Nanodiamonds Via Surface Chemical Functionalization: A DFT Study

Noam Brown and Oded Hod

Department of Chemical Physics, School of chemistry, The Raymond and Beverly Sackler Faculty of Exact Sciences, Tel-Aviv University, Tel-Aviv 69978, Israel

Abstract: The electronic properties of chemically functionalized nanodiamonds are studied using density functional theory calculations. HOMO-LUMO gap and relative stabilities are calculated for different surface functionalization schemes and diamond nanocrystal morphologies. The effects of chemical decoration on the size and nature of the HOMO-LUMO gap of the various systems considered are discussed in detail. We conclude that surface chemical functionalization has the potential to become an accessible route for controlling the electronic properties of nanodiamonds.

Introduction:

Carbon-based nanocrystalline materials may form a diverse array of lattice structures ranging from (quasi-)zero-dimensional nano-dots through (quasi-)one-dimensional nanotubes and nanowires, two dimensional sheets, as well as three dimensional crystal structures. [1, 2, 3, 4] This is a result of the flexibility of the carbon atom in forming four ($sp^3$ hybridization), three ($sp^2$ hybridization), and two ($sp$ hybridization) carbon-carbon bonds. This diversity provides an excellent platform for material engineering allowing for delicate control over their physical and chemical properties and opening the door for fascinating technological applications in many fields including electronics, [5, 6] biology, [7, 8] medicine, [9] and chemical catalysis. [10]

Specifically, nanodiamonds [11] have become a focus of extensive study due to advances in synthesis techniques such as chemical vapor deposition (CVD) and detonation synthesis. [12, 13] It has been shown that nanodiamonds can coexist along with bucky-diamonds (carbon-onions), when annealed at high temperatures or treated by electron beam radiation [14, 15, 16, 17, 18, 19, 20, 21, 22, 23]. Here, the outer surfaces of the diamond nanocrystals undergo a graphitization process, forming a carbon-onion structure, as a transition from $sp^3$ to $sp^2$ type bonds occurs. This graphitization process, which occurs due to dangling bonds appearing at the diamond's outer surfaces, was observed both experimentally [14, 15, 16] and theoretically. [17, 18, 19, 20, 21, 22, 23] The same studies have also shown that hydrogenation of the nanodiamond surfaces prevents such graphitization by the elimination of the dangling bonds thus stabilizing the crystal structure.

Recently, chemical doping of nanodiamonds has been studied extensively aiming to control their physical properties thus enabling the design of nanoscale diamond-based semiconductors [24] and fluorescent biomarkers [7, 8] with desired physical properties. [25, 26, 27, 28] Apart from chemical doping, several experimental studies have addressed the issues of surface functionalization [29] exploring the possibility to enhance solubility in polar organic solvents and reduce aggregation of the nanoparticles in physiological conditions, [30] as well as to create platforms for catalysis and linkage of organic and bio-organic materials. [31, 32] Despite these experimental efforts and the large body of computational work addressing the issue of nanodiamonds chemical doping, to the best of our knowledge, first-principles calculations on surface chemical functionalization of nanodiamonds have been relatively sparse [33].

In the present paper, we present the results of a density functional theory (DFT) study of the electronic properties and the heat of formation of carbon nanodiamonds of three different nanocrystal morphologies and several surface functionalization schemes. We find that both the HOMO-LUMO gap and the heat of formation are affected greatly by the type of chemical functionalization thus opening a new route for designing control schemes for the fabrication of nanodiamonds with desired physical properties.

The paper is organized as follows, in the next section we describe the diamond nanocrystal structures and the functionalization types that are considered. Next, we describe the computational methods and present our main results. This is followed by a discussion and conclusions of the present work.

Nanodiamonds structures and functionalization schemes:

Three different diamond nanocrystals were studied (see Fig. 1): one tetrahedral (marked as ND-I) and two triangular bi-pyramid structures (marked as ND-II and ND-III). The tetrahedral structure (panels (a) and (b) of Fig. 1) is a triangular pyramid cut out of bulk diamond and functionalized at the outer surfaces. Triangular bi-pyramid ND-II (Fig. 1(c)) is obtained by fusing two ND-I tetrahedrons at a given base. Triangular bi-pyramid ND-III (Fig. 1(d)) is obtained by eliminating one atomic layer from one ND-I structure and fusing it to another ND-I structure such that the two pyramids are rotated by $120^0$ with respect to each other. The dangling bonds of each of the nanodiamond considered were first entirely passivated with hydrogen atoms followed by full structural relaxation. The obtained hydrogenated systems then served as reference for several functionalization schemes considered in this work including fluorination, chlorination, nitrogenation, oxidation, and carboxylation. Coordinates of the fully relaxed systems can be found in the supplementary material.

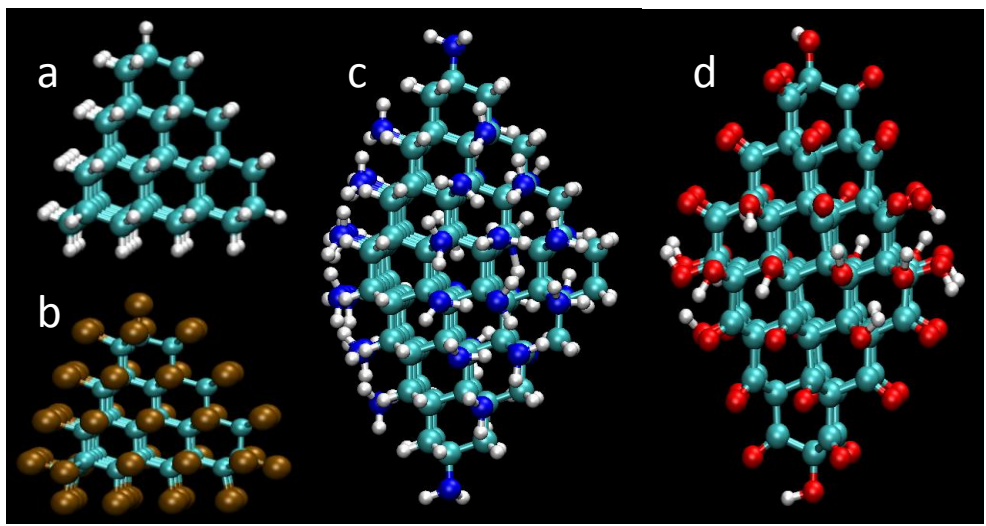

Figure 1: A schematic representation of the different nanodiamond structures considered in this work and some representative functionalization schemes. (a) Fully hydrogenated ND-I. (b) Fully chlorinated ND-I. (c) A partial nitrogenation scheme of ND-II. (d) Fully oxidized ND-III. Cyan, grey, brown, blue, and red balls stand for carbon, hydrogen, chlorine, nitrogen, and oxygen atoms, respectively

ND-I was fully functionalized with fluorine, chlorine, oxygen, carboxylic acid and nitrogen. ND-II was fully hydrogenated and fluorinated, and partially functionalized (30%) with all other substituents considered. Higher coverage schemes of this structure were not considered due to steric hindrance. ND-III was fully functionalized with hydrogen, fluorine, chlorine, oxygen, nitrogen and partially functionalized (25%) with carboxylic acid. When functionalizing the different nanodiamonds with oxygen and nitrogen containing groups, three-fold coordinated carbon atoms (appearing at the lattice surfaces and the apexes) were substituted with a hydroxyl or a primary amine group and carbon atoms having a two-fold coordination (appearing at the lattice edges) were substituted with a ketone or an imine (=NH) group. In order to evaluate the effects of partial functionalization, ND-I was hydrogenated and partially functionalized with chlorine (75%, 30% and 7% coverage); and with oxygen (65%, 30% and 7% coverage). A similar procedure was carried out for ND-III with nitrogen functionalization and oxidation (84%, 25% and 10% coverage).

Computation methods:

All calculations presented have been performed using density functional theory with three different exchange-correlation functional approximations including the local density approximation (LDA), [34, 35, 36] the Perdew-Burke-Ernzerhof (PBE) flavor of the generalized gradient approximation, [37, 38] and the screened-hybrid functional of Heyd, Scuseria and Ernzerhof (HSE), [39, 40, 41, 42] and the double-zeta polarized 6-31G** basis set [43, 44] as implemented in the *Gaussian* 09 suite of programs. [45] Convergence of the results with respect to the choice of basis set was validated with respect to calculations performed at the LDA/cc-pVTZ level of theory producing total energies and band gaps in agreement to within 10%. All considered structures have been fully optimized separately for each functional approximation and basis set used. Presented herein are the results obtained mainly from the calculations performed using the HSE functional and the 6-31G** basis set; this approach has been shown to produce accurate structural and electronic properties of carbon-based nanostructures [46, 47, 48, 49, 50, 51]. Whenever indicated, results of the LDA and PBE calculations can be found in the supplementary material.

HOMO-LUMO gap values reported were extracted from the difference between the highest occupied and lowest unoccupied Kohn-Sham orbitals. Calculations of the relative stabilities of the different structures were performed using the following formula [52, 53]:

(1) $\delta G = E - \chi_C \mu_{diamond} - \sum \chi_i \mu_i$

Here, $E$ - is the cohesive energy per atom of the nanodiamond, $\chi_C$ - is the molar fraction of the carbon atoms, $\mu_{diamond}$ is the chemical potential of the carbon atoms taken as the cohesive energy per atom of bulk diamond calculated at the same level of theory, $\chi_i$ is the molar fraction of the functionalizing atoms of type $i$, and $\mu_i$ - is the chemical potential of the functionalizing atoms taken to be the binding energy per atom of each substituent in its natural form calculated at the same level of theory. For example, the chemical potential of oxygen was calculated as half the binding energy of the triplet state of an $O_2$ molecule. This definition allows for a direct energetic stability comparison between nanodiamonds of different chemical compositions, where negative values represent stable structures with respect to the constituents. It should be noted, however, that this treatment gives a qualitative measure of the relative stability while neglecting thermal and substrate effects and zero point energy corrections.

Results and discussion:
===

**Relative stability**

We start by analyzing the relative stability of the various functionalization schemes of the different nanodiamond structures considered. In Fig. 2, we present the relative stability of the different nanodiamonds, obtained using Eq. 1, as a function of functionalization scheme calculated using the HSE functional approximation (results of the LDA and PBE calculations can be found in the supplementary material). Focusing first on the results for the ND-I system (panel 2 (a)), we find that, compared to the hydrogen passivated systems (leftmost column), full surface fluorination is highly favored. We attribute this to the high electronegativity of the fluorine atoms making them excellent candidates to saturate the surface carbon dangling bonds. While a similar effect could be expected for chlorine atoms, at full coverage, the chlorinated system is found to be less stable than the hydrogenated one by 0.48 eV. This may be attributed to the larger atomic volume of the chlorine atoms leading to enhanced steric effects. This is further supported by the fact that at lower chlorine coverages the relative stabilities of the systems increase monotonously. Partial nitrogenation (30%) results in structural stability comparable to that of the hydrogenated system, whereas partial carboxylation (30%) and oxidation (30%) stabilize the structures by 0.22 eV and 0.20 eV, respectively. Similar trends are obtained for ND-II and ND-III (panels (b) and (c) in Fig. 2, respectively), apart from the case of high nitrogen coverages (100%, 84%) of ND-III where steric effects, again, become important.

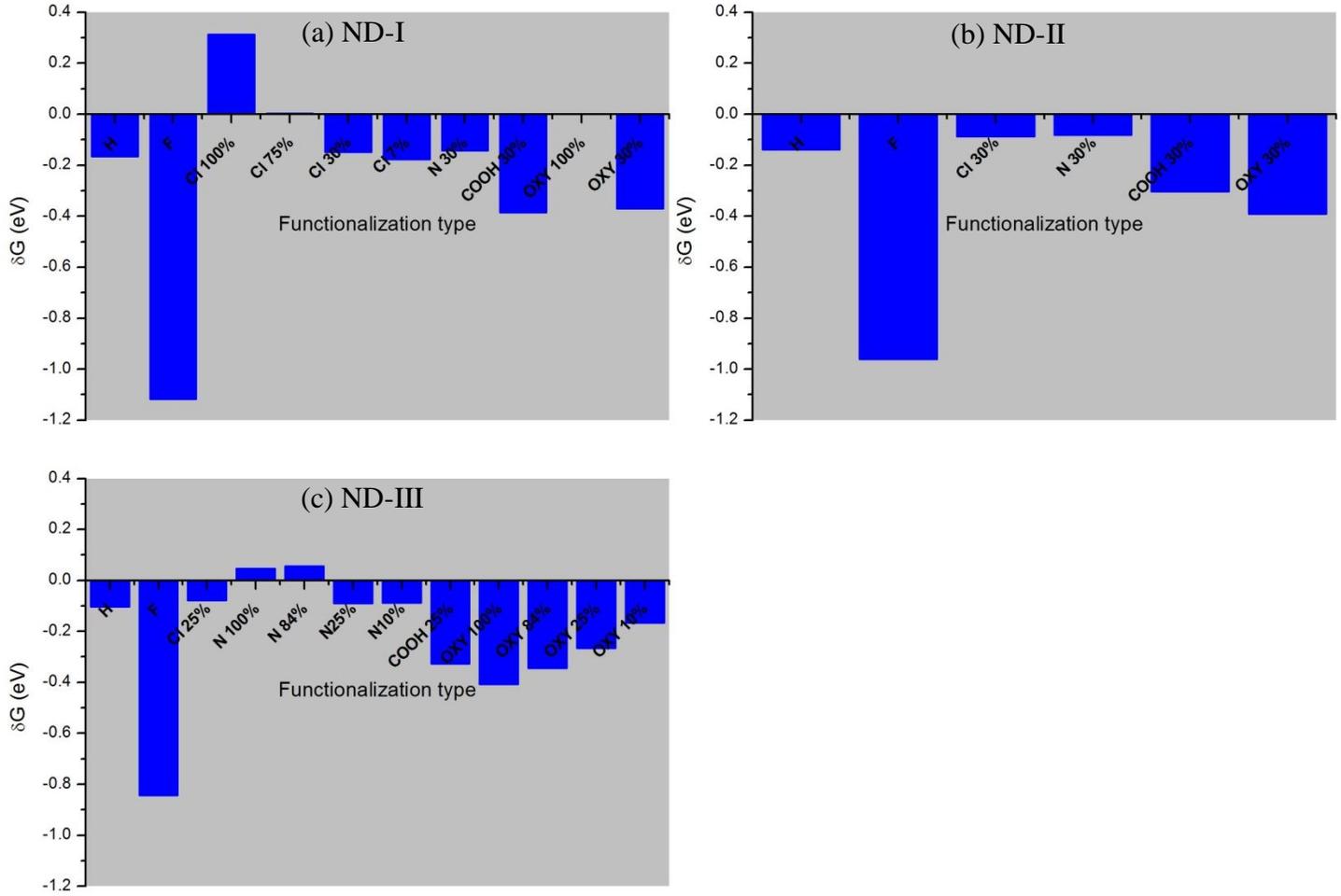

Figure 2: $\delta G$ values, calculated using Eq. (1) at the HSE/6-31G** level of theory, for various surface functionalization schemes of the (a) ND-I, (b) ND-II, and (c) ND-III systems.

We note that most of the functionalization schemes studied produce negative δG values indicating their relative energetic stability. We therefore deduce that under appropriate synthesis conditions diverse functionalization designs are expected to occur, which may lead to the formation of new structures bearing varied physical properties.

**Electronic properties**

After studying the structural stability of the various systems considered we now turn to evaluate the influence of different functionalization schemes on the electronic properties of nanodiamonds. Fig. 3 presents the dependence of the HOMO-LUMO gap on the type of surface functionalization for the three nanodiamond models studied. Concentrating on the HSE results (LDA and PBE results can be found in the supplementary material) all three fully hydrogenated systems possess a high HOMO-LUMO gap of 6.5-7.5 eV. Despite the strong effect of fluorination on the structural stability of the three nanodiamonds, its influence on the electronic structure is relatively minor showing a 3.6% (0.21 eV) gap increase for ND-I, a small gap increase of 2.5% (0.17 eV) for ND-II, and a mild decrease of 5.0% (0.26 eV) for ND-III.

Chlorination, on the other hand, results in a strong reduction of the gap in all three systems. Upon gradual decrease of the chlorine density on the ND-I surface the gap is found to increase. Notably, even at the lowest density considered (7%) the HOMO-LUMO gap is smaller by 11.2% (0.6 eV) than the fully hydrogenated nanodiamond. The effect of partial (25-30%) carboxylation on the ND-I and ND-II HOMO-LUMO gaps is similar to that of partial (30%) chlorination whereas for the ND-III system a smaller gap reduction of 21.3% (0.94 eV) is obtained.

For the nitrogenation functionalization schemes considered we find, again, a relatively small gap decrease of 9.9% (0.74 eV), 16.8% (1.13 eV), and 0.64% (0.04 eV) for low nitrogen coverages (10%-30%) of the ND-I, ND-II, and ND-III systems, respectively. At higher nitrogen coverages (84%-100%) of the ND-III system a considerable HOMO-LUMO gap decrease of up to 54.5% (3.75 eV) is obtained, resulting in a 3.14 eV gap semiconducting nanodiamond. A very similar trend is found for the oxidation scheme considered, where at high coverage (84%-100%) HOMO-LUMO gaps as small as 3.72 eV and 2.76 eV are obtained for the ND-I and ND-III systems, respectively. At lower oxidation coverages (7%-30%) the gap of all three systems considered only slightly reduces from that obtained for the fully hydrogenated system.

(a) ND-I

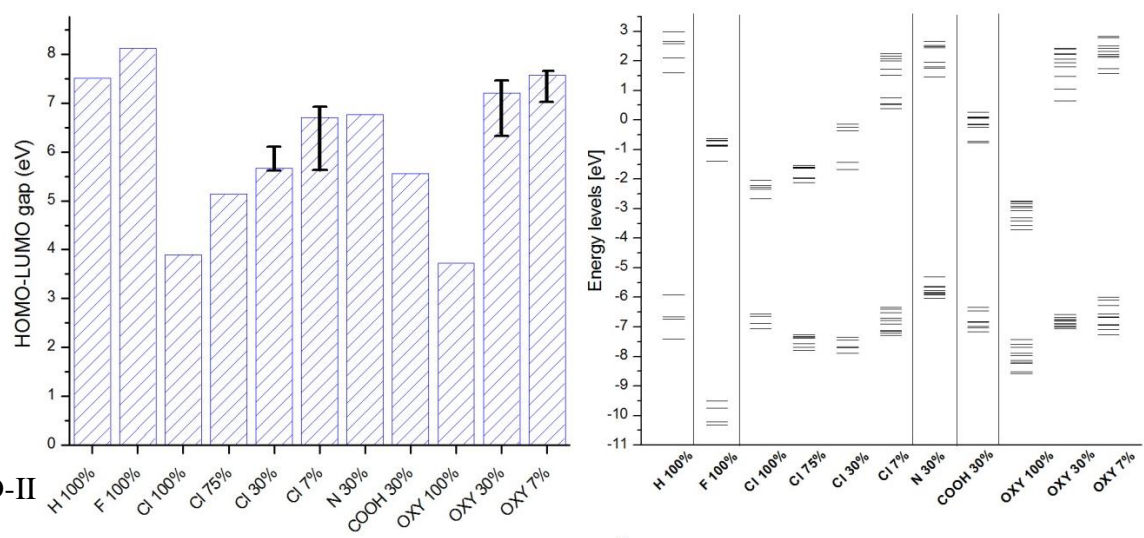

(b) ND-II

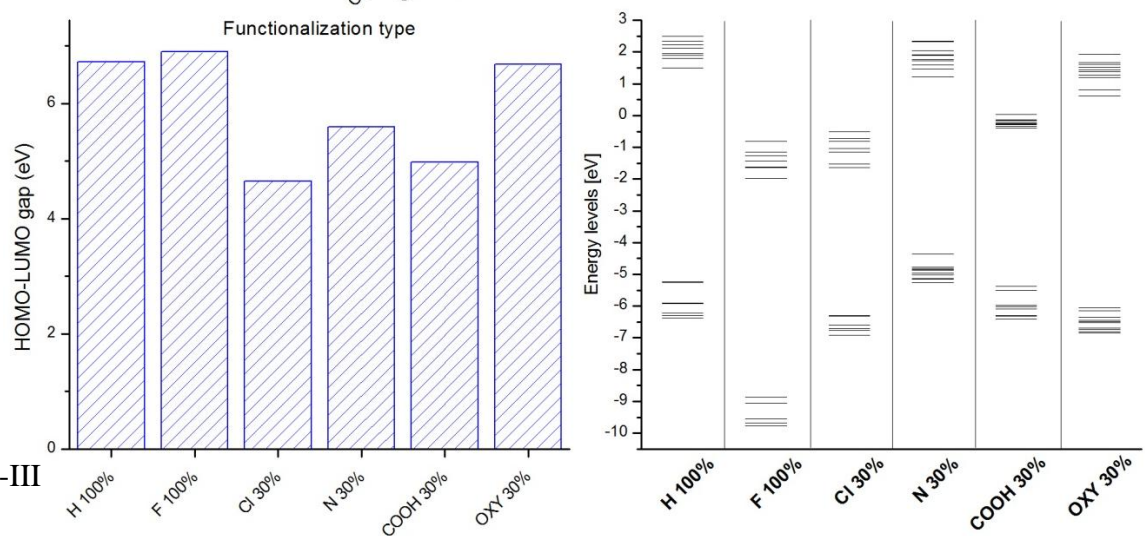

(c) ND-III

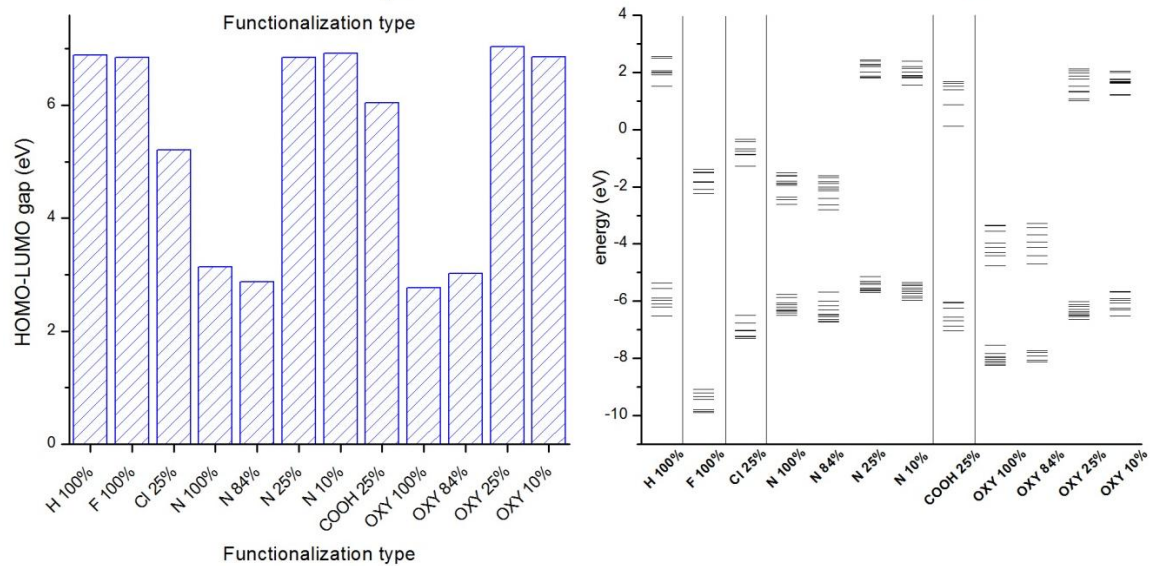

Figure 3: HOMO-LUMO gap vs. functionalization scheme (left panels) and graphical representation of the Kohn-Sham eigenvalues (right panels) for (a) ND-I, (b) ND-II, and (c) ND-III as calculated at the HSE/6-31G** level of theory.

To better understand this behavior we plot, on the right panels of Fig. 3, the energy levels diagrams of the various systems. Focusing on ND-I it can be seen that fluorination results in strong downshift of the whole energy spectrum of the system. Nevertheless, since the HOMO and LUMO Kohn-Sham orbital energies are shifted by a similar amount the overall effect on the HOMO-LUMO gap of the nanodiamond is small. On the contrary, upon full chlorination there is a strong downshift of the unoccupied subspace and a relatively minor effect on the occupied orbitals resulting in the considerable reduction of the HOMO-LUMO gap discussed above. When the chlorine surface coverage in reduced the unoccupied subspace is gradually up-shifted and the HOMO-LUMO gap approaches that of the fully hydrogenated system. A similar trend is obtained for the oxidation scheme where full surface coverage results in strong downshift of the unoccupied subspace with gradual recovery of the hydrogenated system spectrum as the surface coverage reduces. For partial carboxylation, similar behavior is seen where considerable downshift of the unoccupied subspace and a slight down-shift of the occupied orbitals results in the reduction of the HOMO-LUMO gap with respect to the hydrogenated systems. On the contrary, for the partially nitrogenated system a small up-shift of the HOMO orbital energy with almost no change of the LUMO energy leads to the HOMO-LUMO gap reduction. A similar picture arises for the other nanodiamonds considered indicating the general nature of the effect of various decoration schemes on the electronic properties of these systems and supporting our hypothesis that chemical decoration can be used to tune their electronic properties.

Finally, for the mixed decorated structures, we explore the effect of the exact decoration configuration on the relative stability and electronic properties of the system. To this end, we choose three randomly distributed decoration schemes of the 7% and 30% chlorinated and oxidized ND-I structure (see Figs. 4 and 5). Interestingly, we find that, per given atomic surface coverage, the effect of changing the decoration configuration on the relative stability of the system is minor, with changes not exceeding 5%. This may be attributed to the fact that the different decoration configurations, having the same chemical composition, present a very similar chemical bonding scheme. A very different picture arises when considering the HOMO-LUMO gap of the various systems. Here, changing the decoration configuration, while maintaining a fixed coverage, results in considerable gap variations. As can be seen in panel (a) of Fig. 3, for the low-coverage (7%) chlorination scheme the HOMO-LUMO gap can vary by as much as 1.27 eV (which is 22.6% of the total gap). Similarly, for the 30% coverage oxidation scheme the HOMO-LUMO gap may change by ~0.85 eV (11.8% of the total gap). Somewhat smaller variations were obtained for the 30% chlorination and 7% oxidation schemes. To better understand the variations of the HOMO-LUMO gap at a given surface coverage we plot, in Figs. 4 and 5, the energy diagrams of the ND-I system with the various 7% chlorination and 30% oxidation schemes, respectively. As can be seen, the exact decoration scheme has a strong effect both on the density of states in the vicinity of the gap and on the positioning of the HOMO and LUMO energies. This is further reflected by the fact that both HOMO and LUMO orbitals are strongly affected by variations of the decoration scheme (see Fig. 6) as compared to the fully hydrogenated system. Specifically, the LUMO orbitals of the partially decorated systems tend to localize at different apexes where chemical substitution is introduced.

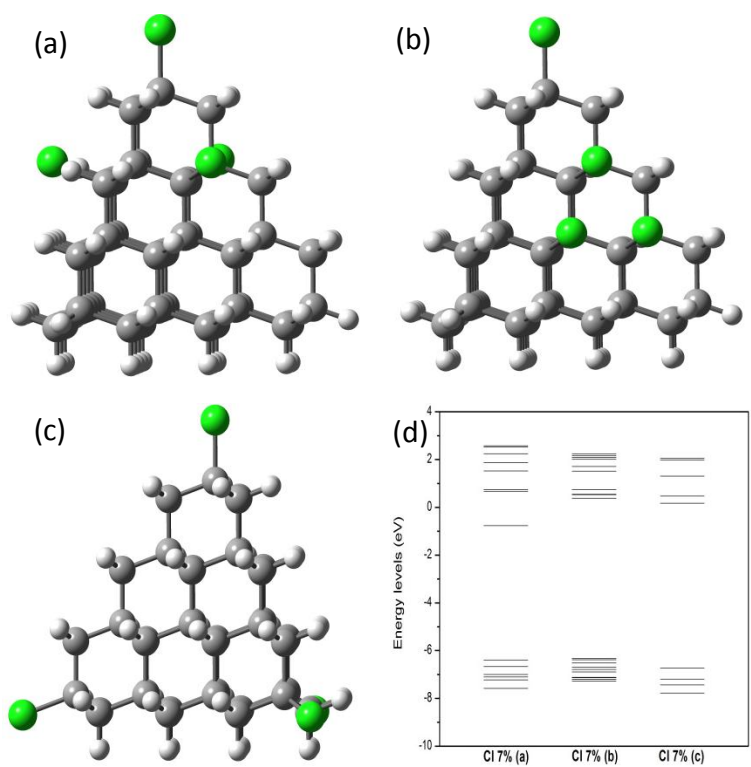

Fig. 4 – decoration schemes (panels (a), (b) and (c)), and Kohn-Sham low-energy eigenvalue spectrum (panel (d)) for the 7% surface chlorinated ND-I system calculated at the HSE/6-31G** level of theory.

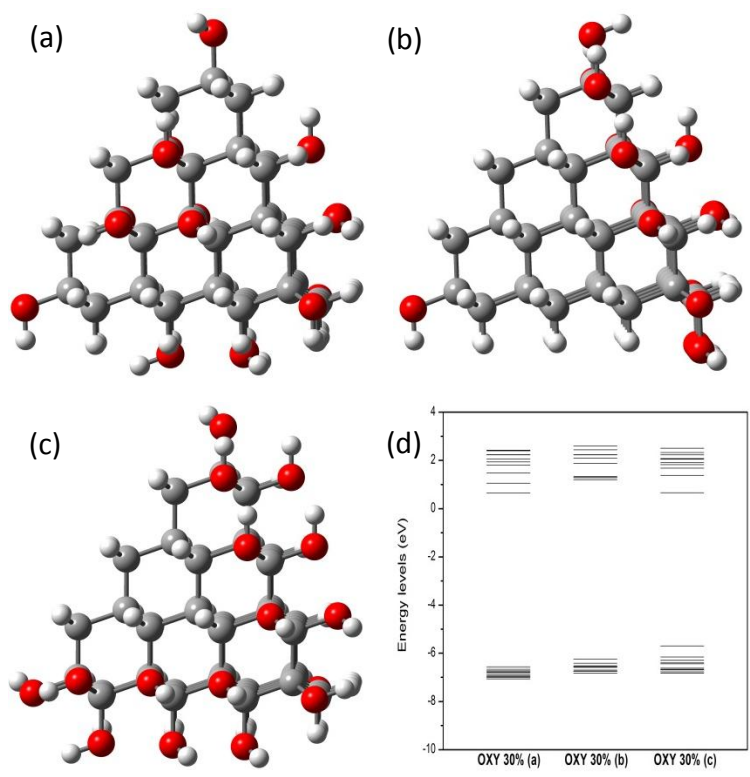

Fig. 5 – decoration schemes (panels (a), (b) and (c)), and Kohn-Sham low-energy eigenvalue spectrum (panel (d)) for the 30% surface oxidized ND-I system calculated at the HSE/6-31G** level of theory.

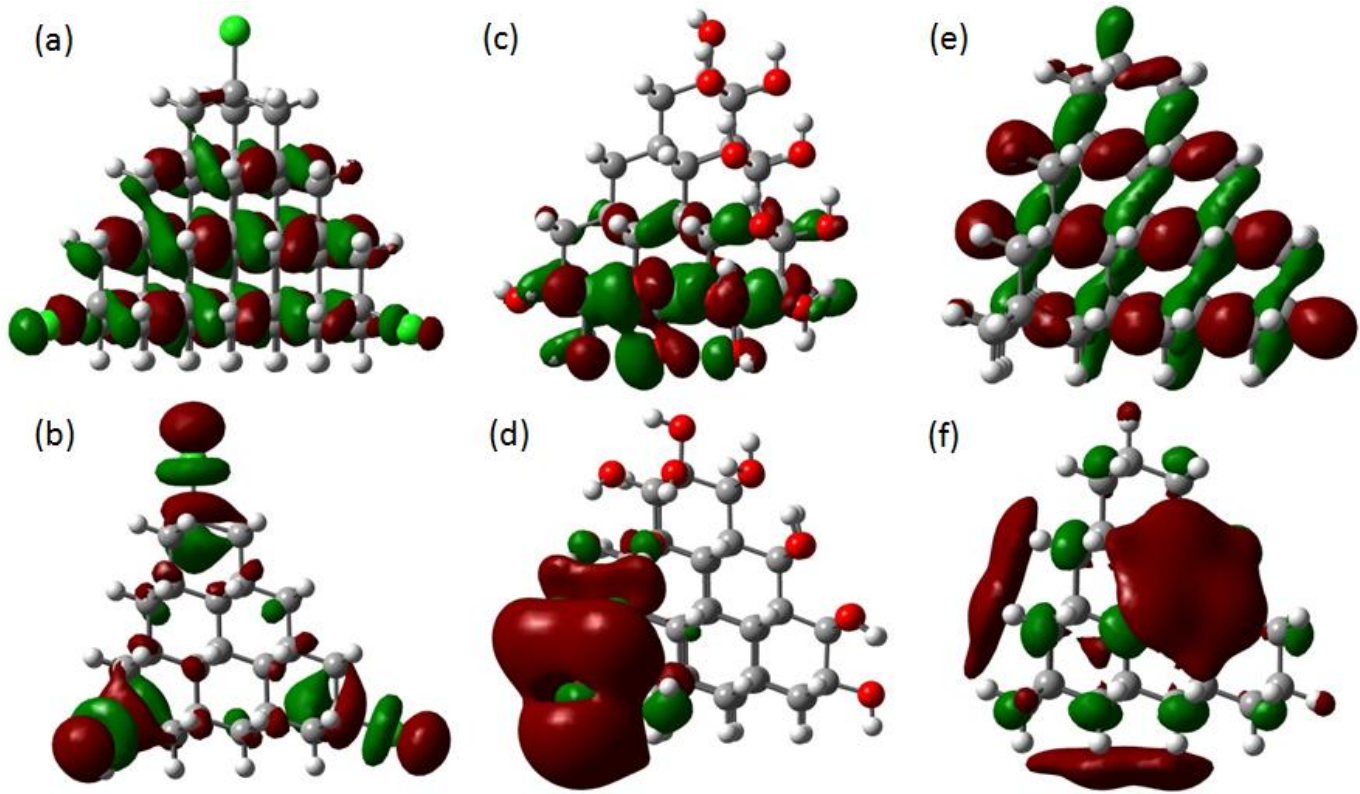

Fig. 6 – representative HOMO (upper panels) and LUMO (lower panels) isosurfaces of chosen 7% chlorination (leftmost panels), 30% oxidation (center panels), and full hydrogenation (rightmost panels) schemes of the ND-I system. Similar results are obtained for the other decoration schemes of the various surface coverages.

## Summary and Conclusions:

In this paper, we have examined the effect of chemical surface functionalization on the structural and electronic properties of nanodiamonds of different crystal morphologies. Various functionalization schemes have been considered including hydrogenation, fluorination, chlorination, oxidation, nitrogenation, and carboxylation. Most chemical surface decoration schemes considered were found to be energetically stable with respect to their chemical constituents. Specifically, surface fluorination was found to considerably enhance the stability of the nanodiamonds resulting in the most favorable structures. Diverse effects of surface chemistry on the electronic structure of the nanodiamonds have been obtained with HOMO-LUMO gaps varying in the range of 2.8-8.1 eV, depending on the morphology of the nanodiamond and its surface functionalization scheme. For the mixed functionalization schemes it was found that the HOMO-LUMO gap varies gradually with the surface coverage of the decorating groups. Furthermore, at a given surface coverage, the gap was found to be sensitive to the specific decoration scheme. These results suggest that surface chemistry may be used as a viable tool for tailoring the structural stability and electronic properties of carbon-based nanodiamonds.

## Acknowledgements

This work was supported by the Israel Science Foundation under Grant 1313/08, the Center for Nanoscience and Nanotechnology at Tel Aviv University, and the Lise Meitner-Minerva Center for Computational Quantum Chemistry.